\documentclass[sn-mathphys]{sn-jnl}
\newcommand{\red}[1]{\textcolor{black}{#1}}

\jyear{2021}%

\begin{document}

\title[Edge spin wave transmission through a vertex  domain wall in triangular dots]{Edge spin wave transmission through a vertex  domain wall in triangular dots}

\author[1]{\fnm{Diego} \sur{Caso}} 

\author[1,2]{\fnm{Farkhad} \sur{Aliev}} 


\affil[1]{\orgdiv{Departamento de F\'isica de la Materia Condensada C03}, \orgname{Universidad Autónoma de Madrid}, \orgaddress{\street{Cantoblanco}, \city{Madrid}, \postcode{28049}, \country{Spain}}}

\affil[2]{ \orgname{Instituto Nicol\'as Cabrera (INC) and  Condensed Matter Physics Institute (IFIMAC), Universidad Aut\'onoma de Madrid}, \orgaddress{\street{Cantoblanco}, \city{Madrid}, \postcode{28049}, \country{Spain}}}



\abstract{Spin waves (SWs), being usually reflected by domain walls (DWs), could also be channeled along them. Edge DWs yield the interesting, and potentially applicable to real devices property of broadband spin wave confinement to the edges of the structure. Here\red{,} we investigate through numerical simulations the propagation of quasi one-dimensional spin waves in  triangle-shaped amorphous YIG ($Y_3Fe_5O_{12}$) micron sized ferromagnets as a function of the angle aperture. The edge spin waves (ESWs) have been propagated over the corner in triangles of 2 microns side with a fixed thickness of 85 nm. Parameters such as superior vertex angle (in the range of 40-75$^\circ$) and applied magnetic field have been optimized in order to obtain a higher transmission coefficient of the ESWs over the triangle vertex. We observed that for a certain aperture angle for which dominated ESW frequency coincides with one of the localized DW modes, the transmission is maximized near one and the phase shift drops to $\pi/2$ indicating resonant transmission of ESWs through the upper corner. We compare the obtained results with existing theoretical models. These results could contribute to the development of novel basic elements for spin wave computing.}

\maketitle



\section{Introduction}\label{sec1}

Despite the growing success of magnonics in recent years, particularly due \red{to} its potential application for spin wave computing and signal processing \cite{Chumak2022}, the currently available spin-wave devices based on ferromagnetic strip lines are \red{restricted} in the frequency span and have none or limited capability in the redirection of SWs \cite{Chumak2017,Chumak2022}. 

Their typical operating frequencies are below few GHz, and the propagation is confined to linear coplanar waveguides. Magnon excitation is usually achieved through \red{the} coupling between the magnetization and the magnetic field due to microwave (mw) currents. The field is however difficult to confine to sub-micrometre volumes. This impedes the use of the traditional inductive coupling methods. Lara et al. \cite{Lara2017} proposed a technologically new approach which could lead to a radical enhancement of the coupling in small magnonic structures, ultimately promising a full integration of the SW devices into CMOS technology.

This radically new pathway is based on the excitation and propagation of a new class of localized quasi-one-dimensional spin waves, the so called Winter magnons (WMs) \cite{Winter1961}. These spin waves are analogous to the displacement waves of strings and could be excited in a wide class of patterned magnetic nanostructures possessing domain walls. Localized WMs have been identified experimentally in different ferromagnetic structures with DWs. Winter magnons have been excited in circular magnetic dots in double vortex states, being localized along DWs connecting vortex cores with half-antivortices \cite{Aliev2011}. The experiments have been supported by numerical modelling and analytical theory. Garcia-Sanchez et al. \cite{Garcia-Sanchez2015} evidenced theoretically and through micromagnetic simulations non-reciprocal channeling of WMs in ultrathin ferromagnetic films along Néel-type DWs arising from a Dzyaloshinskii-Moriya (DMI) interaction.

\red{Other possible application of WMs include spin wave diode \red{\cite{Lan2015}} or SW propagation along DWs using reconfigurable spin-wave nanochannels or spin textures \cite{Wagner2016, Albisetti2018, Hartmann2021}.} Park et al. \cite{Park2020} and Osuna Ruiz et al. \cite{Osuna2019} investigated the propagation of WMs in patterned structures involving both DWs and single vortex states. WMs excited in exchange-coupled ferromagnetic bilayers have been suggested for their implementation in emerging spin wave logic and computational circuits \cite{Sluka2019}.


\red{An entirely different proposal of SW transmission and processing uses Winter magnons confined to edge DWs in ferromagnetic geometries such as triangles or rectangles in configurations created by an in-plane (IP) bias field} \cite{Lara2013,Lara2017}. The resulting state indeed supports the excitation and detection of SWs locally in magnonic logic gates, leading to a natural decrease in device dimensions. Control over edge-localized SWs was accomplished by \red{Zhang} et al. \cite{Zhang2019} by placing magnetic structures adjacently to a propagating microstrip. Additionally, micromagnetic simulations done by Gruszecki et al. \cite{Gruszecki2021} demonstrated that the edge of a structure with locally confined SWs could be used to excite plane waves with twice its frequency and less than half its wavelength.



Our work investigates numerically the excitation, propagation and control of edge spin waves in micron sized YIG triangles with different apertures. We have observed \red{a} resonant enhancement of \red{the} ESW transmission for certain aperture angles accompanied by a $\pi/2 $ phase shift of the propagated spin wave. We link this effect to the interaction of the surface spin waves with the vertex domain wall.

\section{Results and Discussion}\label{sec2}
\subsection{\label{sec:level3}Optimization of the Exchange Energy Channels with Applied Bias Field}

\begin{figure}[tbp]
\begin{center}
\includegraphics[width=0.95\linewidth]{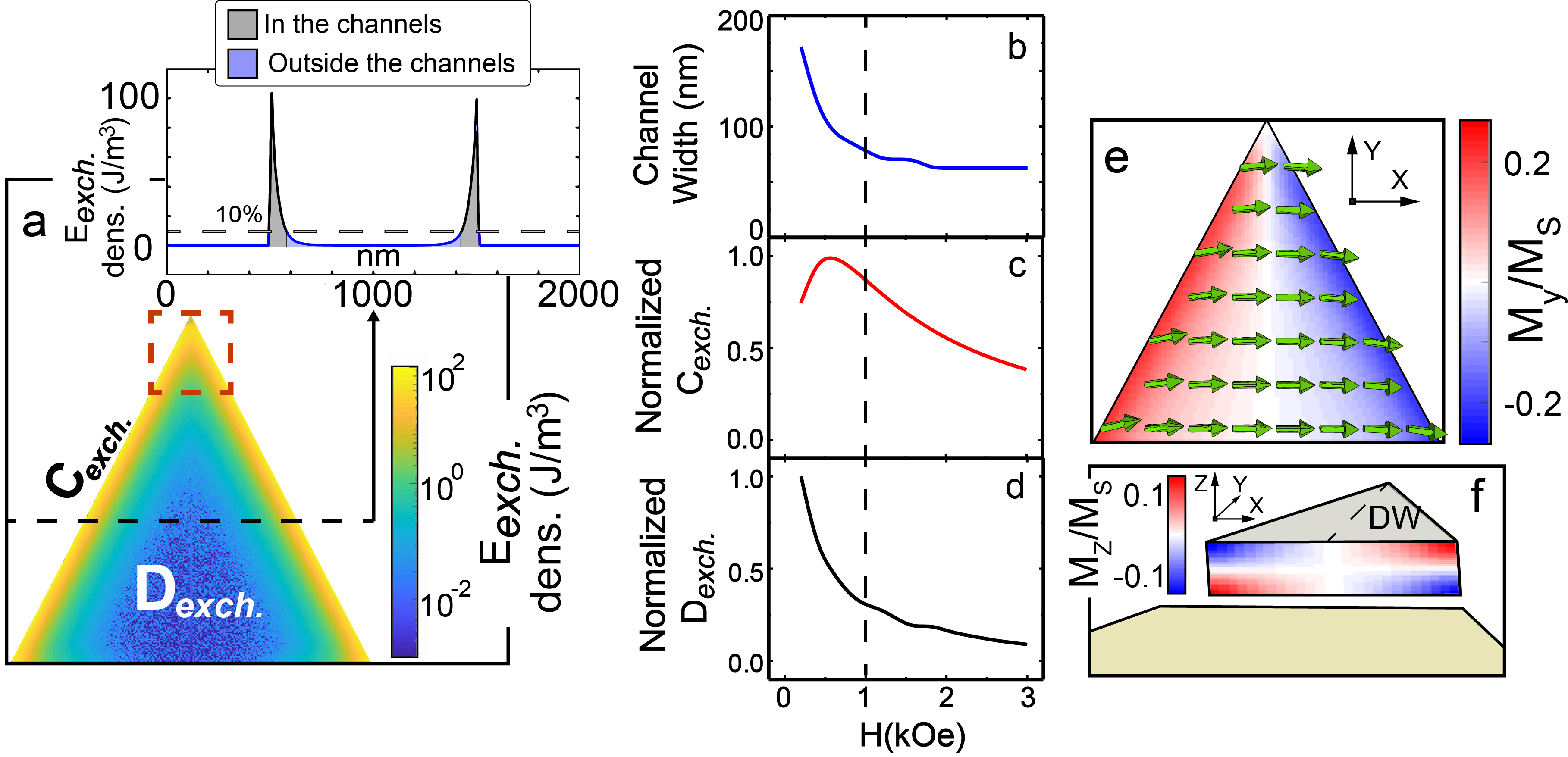}
\caption{(a) Shows the exchange energy density (E$_{exch.}$ dens.) distribution of the YIG triangle with an applied 1 kOe \red{IP} magnetic field parallel to the base of the geometry. \red{Inset shows a cross section of the exchange energy density at the middle of the dot, indicating the way to evaluate the exchange energy channel width.} Part (b) represents the exchange energy channel width vs. the \red{IP} applied field. (c) Indicates how the \red{normalized to maximum value exchange energy in the channel (C$_{exch.}$)} varies with the intensity of the applied field. Part (d) shows the \red{normalized to zero field} delocalized exchange energy (D$_{exch.}$) \red{against the applied field}. \red{Dashed lines in (b), (c) and (d) indicate the optimal applied field (1 kOe) used in the dynamic simulations.} \red{Parts (e) and (f) correspond to the zoomed area surrounded by the red dashed rectangle in (a).} (e) Illustrates \red{the variation of the top corner} static \red{M$_y$ component} magnetization \red{profile in the bulk} normalized by the saturation magnetization (M$_s$). Part (f) reveals an \red{out-of-plane (OOP) profile, limited by the base of the dashed red rectangle in (a), of the M$_z$ magnetization component normalized by M$_s$}. }
\label{exch}
\end{center}
\end{figure}

To get the ground state magnetization distribution a static bias field is applied parallel to the base of the triangle. Once the system is relaxed, the exchange energy distribution will be similar to Fig. \ref{exch}a.  \red{Under a DC field parallel to the base of the triangle, the edge magnetic moments rotate and therefore the  internal magnetic field distribution is minimized near the lateral dot edges. Edge spin waves confined to this potential well can  propagate along the nanostructure edges \cite{Lara2017}}. 

Depending on the strength of the magnetic field, the exchange energy will be accumulated in a larger or lesser \red{degree} in the lateral edges of the triangle, shaping two excess exchange energy edge channels \cite{Lara2017}. \red{The exchange energy channels boundaries are determined here by performing exchange energy cross sections at the middle of the in-plane size of the dot (see inset of Fig. \ref{exch}a), and established where the exchange energy density in the channels is reduced by 90$\%$ from its maxima, determining their width, as presented in Fig. \ref{exch}b). \red{The maximum value of exchange energy density in the channels determines the exchange energy in the channel (C$_{exch.}$), illustrated in Fig. \ref{exch}c, and the delocalized energy (D$_{exch.}$) shown on  Fig. \ref{exch}d is the total summatory of exchange energy outside the channel boundaries. Both quantities are normalized by the maximum value achieved over the applied field.}} \red{The accumulation of exchange energy on the lateral edges of the triangle with magnetic field is a phenomena reflected on} Figs. \ref{exch}b - \ref{exch}d: at small fields the exchange energy \red{density} progressively transfers from being delocalized \red{(exchange energy outside the channel)} to being part of the exchange energy channels. Logically, when stronger field is applied \red{the edge exchange energy channels are being narrowed and there is a better localization for the possible propagation of ESWs} (see Fig. \ref{exch}b). \red{For fields greater than 800 Oe however}, the exchange energy localized in the edges drops, \red{Fig. \ref{exch}c}. This results in a weakening of the channels, which is not desirable for SW propagation. 

Hence, we chose to use 1 kOe applied field for our micromagnetic simulations. \red{The application of this IP} field results in reasonably high exchange energy in the channels for the SW propagation, well localized at the edges of the triangle, \red{as well as reasonably low delocalized exchange energy in the bulk of the system Fig. \ref{exch}d, which implies that the spin waves will less likely travel through the non-edge region of the triangle and cause interferences on the opposite edge from the source}.

Relaxing the magnetic system with no applied field yields in a configuration with an excess of exchange energy as a perpendicular barrier from the middle of the triangles base up to the top corner, which could potentially also be used to propagate SWs. When the 1kOe IP field is applied and the system is relaxed, the remnants of this barrier survive in the top corner. This is related to a domain wall that is originated in the top corner in this particular configuration (see Figs. \ref{exch}e, f).

After the bias field is applied and the system is relaxed, the static magnetization distribution is mostly IP saturated in the direction of the field. However, close to the edges of the triangle the magnetization becomes increasingly parallel to them due to the minimization of stray fields. In the upper corner, this translates as a complete change of the IP magnetization component perpendicular to the base (Y axis) from one side to another of the triangle, leading to a soft 30$^\circ$ Néel-type DW in the top vertex (see Figs. \ref{exch}e, f). However, this \red{statement} is accurate only \red{for} the bulk of the structure, since the magnetization has an increasing out-of-plane (OOP) component in the surfaces of the triangle (see Fig. \ref{exch}f), leading to a state were the DW minimum-energy configuration is intermediate between Bloch and Néel. The resulting state is close to having a weak DMI-like effect \cite{Buijnsters2016}. This topological anomaly has in itself its own magnetic texture and it is of interest to understand spin wave propagation through such structures. The DW spin configuration is head-to-tail, however, extreme high-reflection and low-transmission effects that would occur in 90$^\circ$ Néel-type head-to-tail DWs are negligible due to the softness of the DW, leading to an almost transparent structure in terms of transmission \cite{Hamalainen2018}. The width of the DW is measured at $\sim$ 180 nm.

\subsection{\label{sec:level4}Propagation of the Edge Spin Waves}
\begin{figure}[tbp]
\begin{center}
\includegraphics[width=0.7\linewidth]{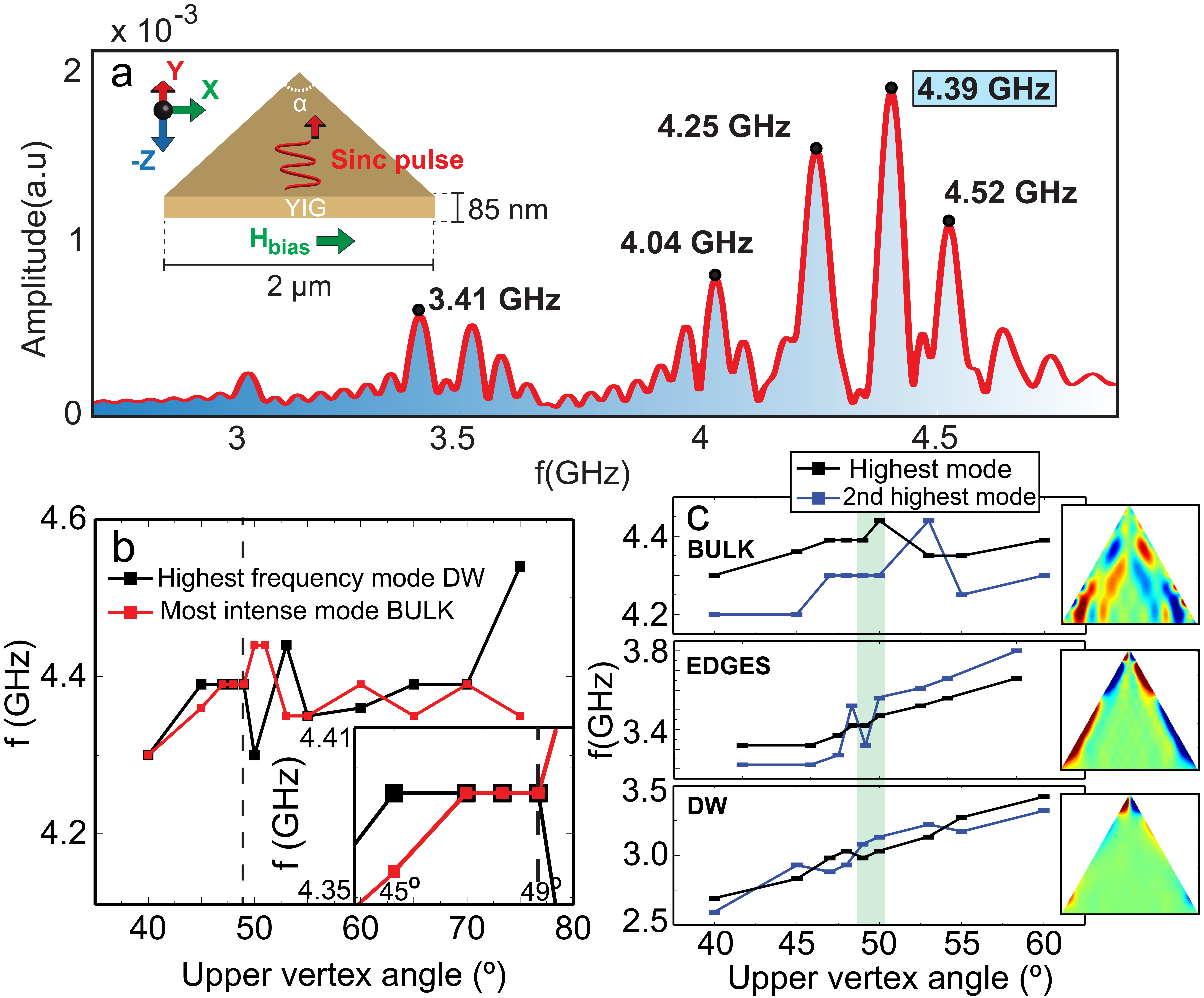}
\caption{(a) Average modes of the whole YIG triangle's OOP magnetization component visualized in the particular case of a 49$^\circ$ aperture. The most intense mode (in this case 4.39 GHz) is to be excited in the left corner of the triangle. Inset shows a sketch of magnetic field distribution in the system: a 1 kOe DC field parallel to the base of the triangle and a smaller AC pulse perpendicular to it. (b) Frequency of the mode with the highest frequency in the DW and the most intense bulk mode \red{(corresponding to the posterior SWs excitation frequency)}  against the angle aperture. For a wide range of angles in the high transmission region or fairly close to it, some DW modes frequencies agree with bulk modes, i.e, when exciting with this frequency they couple and result on a resonant system, which amplifies the transmission through the DW. \red{Inset of (b) is an enhancement in the high transmission regime, showing the overlapping of modes. (c) Two most prominent modes in the bulk, edges and DW against the corner aperture from 40 to 60 degrees. In light green is highlighted the high transmission regime. Inset for each graph in (c) corresponds to a typical profile of the most intense mode for the bulk, edges and DW. }}
\label{pulse49}
\end{center}
\end{figure}

 The generated DW has its own associated eigenmodes that can be dynamically stimulated, this is key to understand SW transmission and further effects that will be discussed later on. After a 20 ns sinc-shaped pulse (see Methods), we found that the response of the whole system to the pulse resulted in clear distinguishable eigenmodes (bulk modes from now on). For all of the analyzed angles, from 40 to 75 degrees, the observed eigenmodes were restricted in the 3 GHz to 5 GHz range (see Fig. \ref{pulse49}a for the modes of a 49$^\circ$ triangle). The mode of the highest amplitude, however, is closer to 4.4 GHz, slightly oscillating for the different triangle apertures (Fig. \ref{pulse49}b). 

\begin{figure}[tbp]
\begin{center}
\includegraphics[width=1\linewidth]{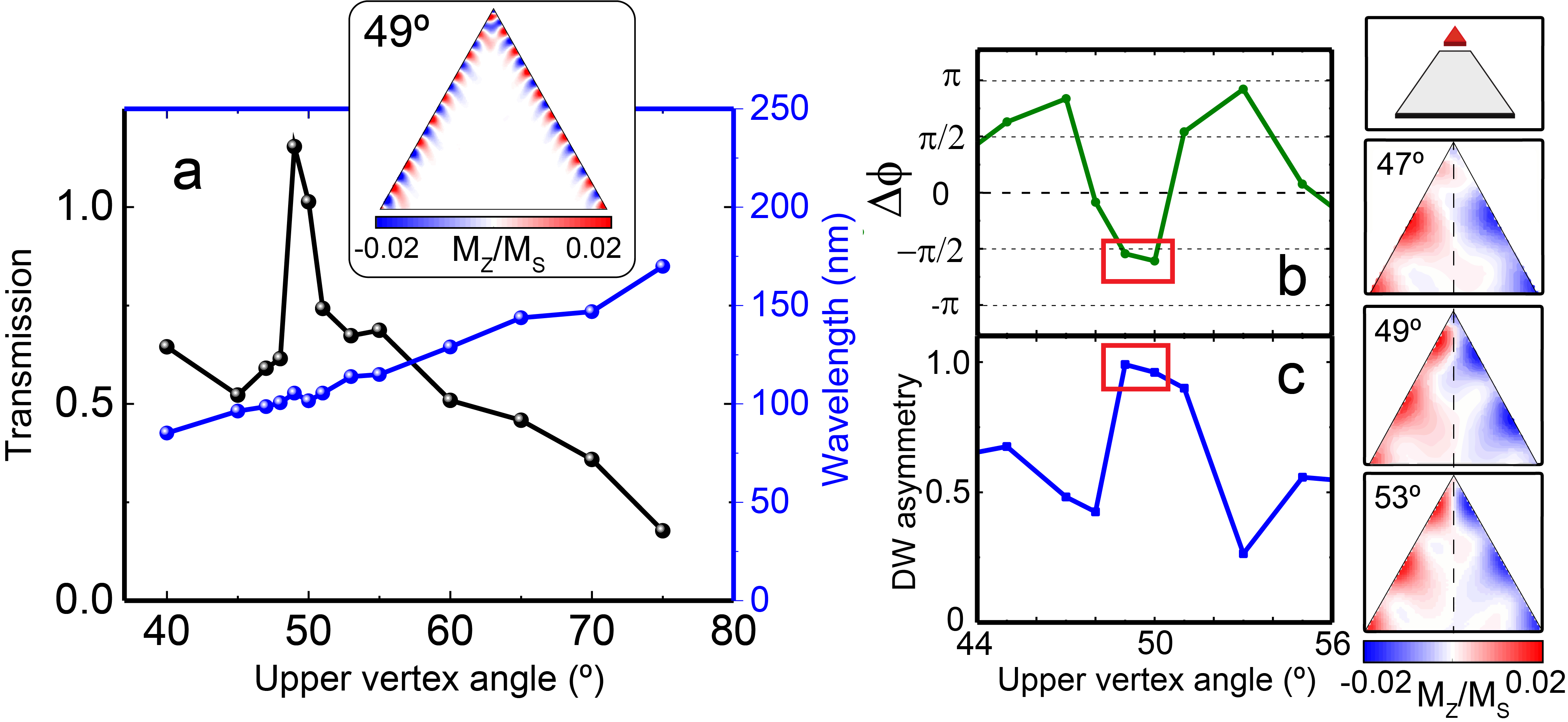}
\caption{(a) Transmission coefficient \red{and wavelength} of the ESW for an upper vertex angle aperture between 40 and 75 degrees analyzed in the bulk of the triangle. The distinct transmission peak at 49-50$^\circ$ indicates that the spin wave propagates almost perfectly through the edges of the triangle. Inset shows the ESW propagating through the edges a for 49 degree angle \red{dot}. The excited frequencies correspond in each case \red{to} the most intense bulk modes. (b) Phase shift of the ESW induced at the upper vertex DW. At the 49-50$^\circ$ aperture angle, which is the high transmission range, the ESW experiences a $\pi$/2 phase shift, \red{indicating} the possible existence of resonance. (c) \red{Normalized} DW asymmetry between the two lateral sides of the triangle in the propagation of the ESW indicates a more asymmetrical propagation for the high transmission angles. Insets reveal an enhancement of the upper vertex DW, \red{showing a snapshot of the SWs once the propagation is steady} for three different angle apertures. \red{Enhanced area is marked in red in the top inset, which constitutes approximately only 17.5$\%$ of the in-plane length of the whole dot.}}
\label{sw}
\end{center}
\end{figure}
The analysis of the eigenfrequencies is also done for local known specific magnetic structures such as the edges or the upper vertex DW. This allows to differentiate the propagation of the spin waves through three different magnetic structures with their own modes: bulk, edges, and the DW. Local analysis of the eigenmodes displays that generally f(bulk) $>$ f(edges) $>$ f(DW) (f being the frequency of the modes) \red{(see Fig \ref{pulse49}c)}. However, these modes are not overwhelmingly separated (all of them are between 1 and 5 GHz), and one can exploit this fact to excite the system at matching frequencies between two -or more- of these structures, resulting in a resonance-like system in which energy is being pumped from the magnetic structures to the spin wave to \red{boost and perpetuate the propagation}. 

Local excitation of the spin waves is done at the left corner of the triangle \red{(see Methods)}, where the magnetization is conflicted between being parallel to the left edge or to the base, which is the bulk magnetization direction. 
\red{Our micromagnetic simulations indicate that it is possible to excite edge spin waves propagating either from the left or the right vertex with a difference in transmission below 5$\%$ and a difference in the excited wavelength below 2$\%$. Interestingly, the amplitude of the propagated SW was found to be about twice larger when ESWs are excited from the left vertex (configuration discussed in this manuscript). The ESW intensity difference could be due to the difference in the angle between the mw excitation (directed perpendicularly to the triangle side) and the direction of the static local magnetization in left and right corners.}

The used \red{frequencies} for the mw excitations are the most intense detected bulk modes \red{(red line in Fig. \ref{pulse49}b)}. However, these excitations are directed effectively perpendicular to the left edge of the triangle. Thanks to this and to the exchange energy channels, we can "trick" the spin wave to being almost completely localized in the edges of the system (see inset of Fig. \ref{sw}a), even though the excited modes are present in the whole system. 

Spin waves propagated from the left to the right corner of the triangle (see Supplementary videos) show a range of aperture angles in which there is a peak in transmission \red{(see Methods section for a description of the transmission analysis)}, even slightly surpassing the value of 1 for 49$^\circ$, which means the edge localized spin wave is \red{a bit} more intense in the right side of the triangle \red{than on the left one, where the source is placed} (see Fig. \ref{sw}a). 
\red{We tentatively explain the transmission \red{coefficient exceeding one} obtained for the aperture angle of 49 degrees, as a result of ESWs and DW modes excited through their resonant interaction with bulk modes. This is backed up by the fact that at the high transmission regime angles some of the highest frequency DW modes coincide with the most intense main modes of the whole system (Fig. \ref{pulse49}b), probably providing an enhanced excitation of the upper vertex DW by the delocalized SWs and therefore effectively boosting the ESWs amplitude at the right side of the triangle.}

\red{To verify the possible resonant transmission for specific frequencies we performed a simulation of a 49 degree triangle excited at an arbitrary frequency of 4.35 GHz (varying less than a 2$\%$ the frequency of the resonant mode), which does not correspond to any mode in the system. Although the ESW signal is propagating (probably because the excitation is close in frequency to other modes), the resulting transmission dropped from about 1, obtained for the most intense mode of 4.39 GHz, to 0.36.}

The \red{strong} dependence in ESW transmission on the aperture angle remarks the importance of the upper vertex DW topology in the transmission process. \red{As we mentioned, an increment in transmission of the SWs in a narrow range of top corner aperture angle could be due to the concordance between the most intense bulk SW modes and the highest frequency DW modes} (see Fig. \ref{pulse49}b). This correlation is difficult to disregard, and backs up the expected result in a resonant system with two sources. A deep analysis of the process in the upper vertex DW when the SW is propagated helps to understand these effects: for the high transmission angle range, the SWs experiments a phase shift \red{(see Methods)} close to $\pi$/2 (marked in red in Fig. \ref{sw}b), which is in agreement with some mechanical systems in resonance and previous recorded phase shifts in Néel-type DWs \cite{Wojewoda2020}. 

The calculated DW asymmetry \red{(see Methods section)} along an axis parallel to the Y axis that divides the upper vertex in two (see inset of Fig. \ref{sw}c) is also maximum for the angles of high transmission coefficient, which is highly correlated to the $\pi$/2 phase shift (Fig. \ref{sw}c). At 49$^\circ$ (see inset of Fig. \ref{sw}c), the incoming and departing signals at the corner \red{do not interfere with} each another (which happens due to the $\pi$/2 phase shift), showing the high efficiency of the local DW SW propagation for this angles. However, for an angle outside of the high transmission range this \red{statement} is not true \red{since} the incoming and departing SWs \red{collide at the corner}, resulting in a transmission not as efficient.

\subsubsection{\label{sec:level5a}Dispersion Relations}
\begin{figure}[tbp]
\begin{center}
\includegraphics[width=0.65\linewidth]{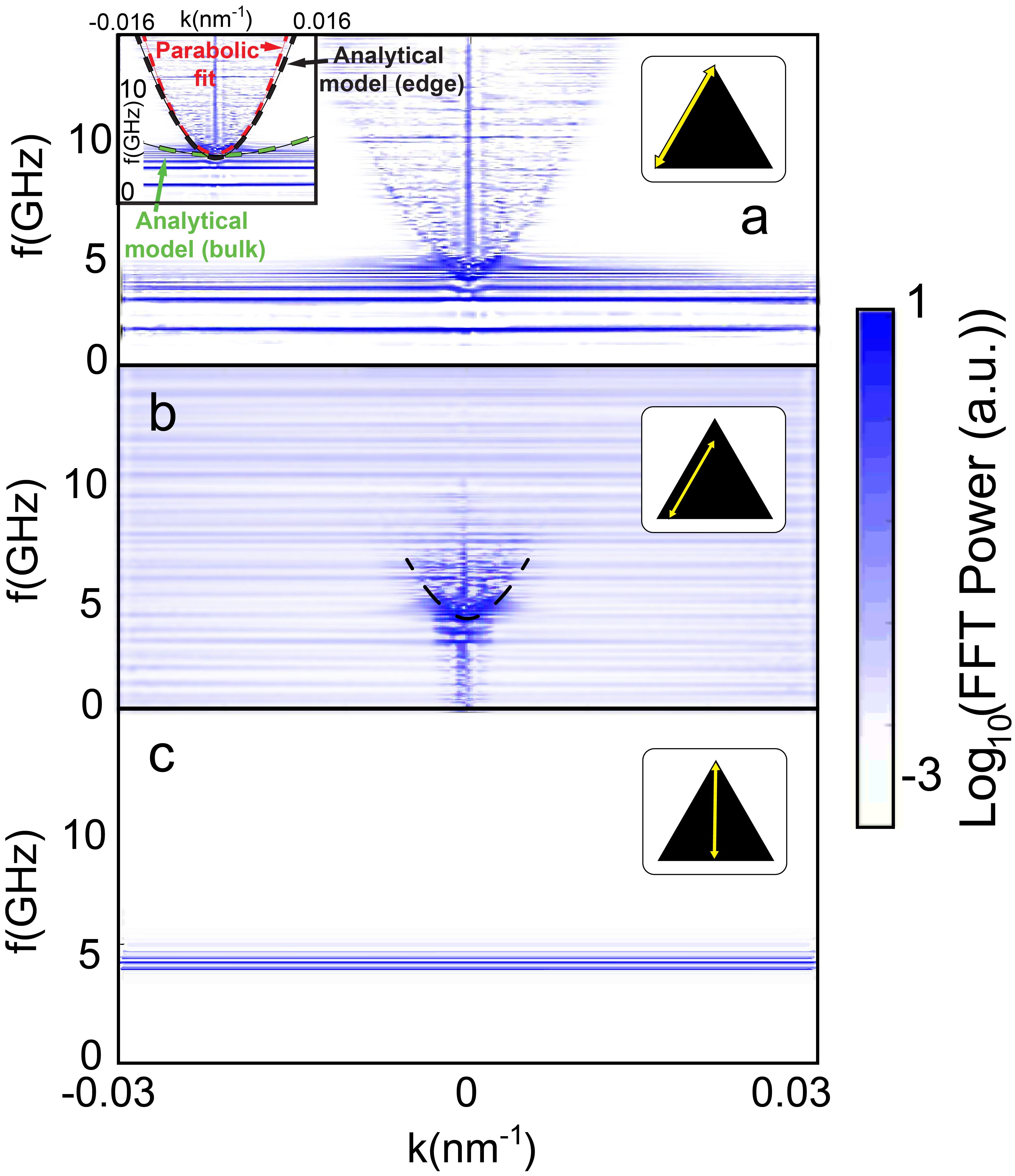}
\caption{Dispersion relation in a 49 degree angle aperture triangle for (a) the left edge of the triangle from the left vertex all the way to the top. Both dynamical regimes are represented in the chosen path: the low-frequency modes constant in frequency and the parabolic ones, at higher frequencies. (b) The left side of the dot with a separation of $\sim80 nm$ from the edge. Although noticeable, both mode regimes are less intense than in the case of the left edge analysis. \red{The black dashed line indicates the less intense parabolic dispersion.}  (c) A straight line perpendicular to the base of the dot, separating it in two halves. In this particular case only the low-frequency modes are present. These results reveal a direct correlation between the low frequency modes and the top corner of the triangle, associated with the DW, and the parabolic mode with the edge of the system}
\label{dispersion}
\end{center}
\end{figure}

Analyzed SW dispersion \red{(see Methods)} reveals two different dynamic regimes: low-frequency modes, which are constant in frequency along all wavenumbers, and a parabolic mode at higher frequencies (see Fig. \ref{dispersion}a). These regimes are both present when the dispersion is represented from the data obtained in the edge region. Different performed tests reveal that the intensity of the dynamic regimes is dependent on the path chosen for the 2D Fourier Transform analysis, which generates the dispersion relations: a path that crosses through the left edge of the structure is associated with the two dynamic regimes being present (Fig. \ref{dispersion}a). However, they become less intense if the path is separated from the edges (see Fig. \ref{dispersion}b). If the path crosses the triangle from the base all the way to the upper vertex the dispersion relation only shows the constant frequency modes through all wavenumbers (see Fig. \ref{dispersion}c). \red{These frequency constant modes (coinciding with some of the modes presented in Fig. \ref{pulse49}a) most probably originate from the localized DW modes along the analyzed vertical line (remarked by a dashed line in Fig. \ref{exch}f) in contrast with delocalized parabolic-type modes linked to edge spin waves}. Here, the positive k vector branch of the parabolic mode corresponds to the ESWs emitted by the left vertex while the negative k branch describes ESWs emitted by the top corner in direction to the left vertex. In this configuration, these two branches have an equivalent intensity, meaning that the propagation of ESWs is \red{in both directions is comparable}. 

Their approximately parabolic dispersion relation also corroborates the predominantly 1D character of SWs along the edge with an IP field parallel to the base. Indeed, the exact solution for SDWs in a 1D ferromagnetic chain predicts a parabolic k dispersion \cite{Gruner1994}, valid except in the region of k → 0, where the dipolar contribution dominates. The parabolic nature of the dispersion relation of the ESWs also matches the expected for the analytical model proposed by Lara et al. \cite{Lara2017} \red{(see Methods)} if the exchange length in the edges is more (an order of magnitude) than the expected value at the bulk (see inset of Fig. \ref{dispersion}a). This is reasonable since the exchange energy is predominantly accumulated in the edges, which would generate a stronger coupling of the magnetic moments in this region, and thus, a longer coupling distance.

\subsection{\red{Edge Spin Waves Phase Shift at the Vertex Domain Wall }}
Several studies have previously considered theoretically spin wave propagation through a domain wall. Hertel et al. \cite{Hertel2004} predicted that there is a proportionality between the SW phase shifted produced by DW and the angle by which the magnetization rotates inside this domain wall ($\Delta \phi = \Delta$f/2), meaning that the phase of spin waves with k-vector perpendicular to a domain wall changes by a factor of $\pi$/2 when the wave propagates in a ferromagnetic layer through a transverse wall. The changes of phase shift of the spin wave with upper vertex aperture angle observed here suggests a possible topological variability of the DW magnetic texture with angle aperture once the excitation reaches a steady state, as confirmed by our results. \red{The change in propagation direction near the vertex could also be a potential cause of the background variation in phase and amplitude of the transmitted edge waves outside the region where resonant enhancement of the transmission and abrupt change in the phase shift is observed.} However, the model \cite{Hertel2004} also suggests that at the high transmission range, in which the phase shift is almost exactly $\pi$/2, the DW becomes completely transverse. This scenario must be discarded due to the analysis done at the DW, which does not show a situation with a particularly hard DW, even when the spin wave is stabilized. 

On the other hand, Bayer et al. \cite{Bloch2005} predicted analytically that the spin-wave transport through an infinitely extended one-dimensional Bloch-type domain wall induces a finite phase shift without reflection. We deduce then that the slight OOP component in the DW plays a role in the transmission of the wave. Indeed, when the excitation stabilizes, the spin wave propagating on each side of the dot have \red{opposing} OOP \red{magnetization} component, meaning that is key for the energy minimization in the system. Further studies involving more precise control over \red{the} magnetic texture of the vertex domain wall, for example using \red{an} underlying material with spin orbit coupling (\red{such as} a $Pt$ underlayer) are needed to explore the direct link between the vertex DW internal magnetic texture and the spin wave transmission through it.

\section{Conclusions}\label{sec13}

Detailed investigation \red{on} the influence of the upper vertex aperture on transmission and dephasing of edge spin waves in amorphous YIG ferromagnetic triangles shows the possibility of fine tuning of the edge spin wave transmission between two remote corners. The results suggest that \red{an aperture angle of} 49-50$^\circ$ triggers a high transmission response of the spin wave propagation system. Local analysis of the eigenmodes reveals \red{the} following relation: f(bulk) $>$ f(edges) $>$ f(DW) for the vast majority of those SW modes branches. In some instances, however, the local modes \red{seemingly} overlap each other, resulting in an energy pumping into the upper vertex DW structure. The maximum DW asymmetry is found at the high transmission range, as well as a phase shift of $\pi$/2, characteristic of resonant systems and Néel-type DWs. The analysis of the SW dispersion relations reveal that the DW-related modes are constant in frequency, whereas the ESWs modes have a parabolic behavior. The best fit to the analytical model \cite{Lara2017} suggests a possible increase of the exchange length along the edge DW. We conclude that the high transmission mechanism is greatly due to the specific DW topology on each angle aperture, which could indeed raise resonance-like interactions between local and bulk SW modes in the triangular ferromagnetic structure. We believe that this work could contribute to better understand the SW propagation through topological objects and/or magnetic textures.

\backmatter



\bmhead*{Methods} 
The micromagnetic simulations were carried out using the MuMax3 code \cite{Mumax2014}. The typical YIG parameters were used: saturation magnetization M$_s$ = 130 kA/m, exchange stiffness constant A$_{ex}$ = 3.5 $\times$ 10$^{-12}$ J/m and damping constant $\alpha$ = 2.8 $\times$ 10$^{-3}$. The base of the used triangles was 2 $\mu$m and its thickness 85 nm. Discretization was set at 7.8 $\times$ 6.7 $\times$ 4.2 nm per cell (256 $\times$ 256 $\times$ 20 cells) for a standard 60$^\circ$ angle triangle, enough to precisely characterize the upper vertex DW. No anisotropies were added to our structure. To find the eigenfrequencies of a magnetic system, a 2 Oe, 20 ns, sinc-shaped IP magnetic field pulse was applied uniformly perpendicularly to both the base of the triangle and the direction of the applied static field. After relaxing the system, the spin wave eigenmode spectrum is obtained using the Fourier Transform of the OOP component of the magnetization. Knowing the eigenfrequencies, which appear as peaks in the absorption spectrum, a local excitation at a single eigenfrequency can be applied \red{locally} to observe the response of the magnetic \red{system} (i.e., to observe the propagation of spin waves away from the source), for as long as it may be necessary for the spin waves to reach a target area. \red{The excitation source was localized in the left vertex of the triangle, in a small volume of just three cells over all the thickness of the triangle (in-plane area is 256$\times$256 cells$^2$). MW excitation is also directed perpendicularly to the left edge of the structure to ease the propagation and help localize the SW into the edges.}

\red{To characterize the DW asymmetry we first reflect the propagation of one side into the other of the DW, thus creating a map in which completely symmetric SWs maxima or minima cancel each other. Then, we estimate the DW asymmetry as the resulting magnetization in the reflected map as follows: DW asymmetry = $\sum \mid M_z \mid $, the summatory being over all the cells in the map.} 

Dispersion relations are calculated from the simulations using a 2D Fourier Transform of the \red{OOP} magnetization \red{component} along a desired path \red{after the sinc-shaped pulse is applied. The proceeding involves using the output magnetization files from the pulse simulation to record a 1D path of magnetization for each assigned time (keeping only the magnetization from the path that has been chosen). An n$\times$m matrix should emerge from this, where n is the number of cells in the path and m the number of time points in the pulse simulation. Then, the 2D Fourier transform is used to switch it into the reciprocal space, which is what is presented in Fig. \ref{dispersion}.} 

\red{Since the two signals from left and right edge-localized propagated spin waves share one fundamental frequency, its phase difference can be analyzed the same way as two AC signals: we first remove the DC part, i.e, the offset. Then we perform a Fourier Transform on both signals. Since the returned values of the Fourier Transform are in terms of magnitude and phase, the phase angle of each signal can be extracted numerically. The phase shift is calculated in the manuscript as the phase encountered in the left edge subtracted from the phase at the right edge of the dot once the signal reaches equilibrium.}

\red{Transmission has been computed by analyzing the signals of the ESWs (once the steady state in the propagation has been reached) by performing a Fourier Transform of the edge spin wave signals on both sides of the dot (approximately 15 nanometers away from the edge) and determining the amplitude of the Fourier Transform peaks quotient. This method allows the almost total elimination of interferences in the process of analysis that would be characterized by undesirable frequencies in the FT.}

\red{The analytical fit of the ESWs dispersion relation in Fig. \ref{dispersion}a was supported by the model proposed by Lara et al. \cite{Lara2017} for edge localized spin waves in magnetic dots, valid only when the dot's aspect ratio is (thickness/in-plane size) $< <$ 1:
\begin{equation}
\omega^2(\kappa) = \omega_M^2[1+l_e^2\kappa^2+(1-\nu^2)h][l_e^2\kappa^2 + (1-\nu^2)h]
\end{equation}
Where $\kappa$ is the wavevector along the edge of the dot, $h = H/4\pi M_s$ corresponds to the reduced bias magnetic field parallel to the base of the triangle,  $l_e = \sqrt{A/2\pi M_s^2}$ is the exchange length, and $\nu$, which is a parameter of the model that quantifies the ratio of spatial decays between dynamic and static magnetizations, has to be less than one to assure that the frequency of the SWs increases with decreasing h.}

\bmhead*{Acknowledgments}
Authors acknowledge Ahmad Awad, César González-Ruano and Antonio Lara for discussions. The work in Madrid was supported by Spanish Ministerio de Ciencia (RTI2018-095303-B-C55) and Consejería de Educación e Investigación de la Comunidad de Madrid (NANOMAGCOST-CM Ref. P2018/NMT-4321) Grants. FGA acknowledges financial support from the Spanish Ministry of Science and Innovation, through the "María de Maeztu" Program for Units of Excellence in $R\&D$ (CEX2018-000805-M) and "Acción financiada por la Comunidad de Madrid en el marco del convenio plurianual con la Universidad Autónoma de Madrid en Línea 3: Excelencia para el Profesorado Universitario". D.C. has been supported by Comunidad de Madrid by contract through Consejería de Ciencia, Universidades e Investigación y Fondo Social Europeo (PEJ-2018-AI/IND-10364)





\end{document}